# Suris tetrons: possible spectroscopic evidence for four-particle optical excitations of the 2D electron gas

A.V. Koudinov[1,2], C. Kehl[3], A.V. Rodina[1], J. Geurts[3], D. Wolverson[4], G. Karczewski[5]

[1]A.F. Ioffe Physico-Technical Institute of RAS, 194021 St.-Petersburg, Russia
[2]Spin Optics Laboratory, St.-Petersburg State University, 198504 St.-Petersburg, Russia
[3]Physikalisches Institut (EP3), Universität Würzburg, 97074 Würzburg, Germany
[4]Department of Physics, University of Bath, Bath, BA2 7AY, UK
[5]Institute of Physics, Polish Academy of Sciences, 02-668 Warsaw, Poland

The excitations of a two-dimensional electron gas in quantum wells with intermediate carrier density ($n_e \sim 10^{11}$ cm$^{-2}$), i.e., between the exciton-trion- and the Fermi-Sea range, are so far poorly understood. We report on an approach to bridge this gap by a magneto-photoluminescence study of modulation-doped (Cd,Mn)Te quantum well structures. Employing their enhanced spin splitting, we analyzed the characteristic magnetic-field behavior of the individual photoluminescence features. Based on these results and earlier findings by other authors, we present a new approach for understanding the optical transitions at intermediate densities in terms of four-particle excitations, the Suris tetrons, which were up to now only predicted theoretically. All characteristic photoluminescence features are attributed to emission from these quasi-particles when attaining different final states.

The intriguing physics of two-dimensional electron gases (2DEGs) in semiconductor quantum wells (QWs) has been subject of intense study since the 1980s. The systematic tunability of the carrier concentration gives access to a wealth of phenomena for 2DEGs with metallic or non-metallic behavior. For high-concentration 2DEGs ($n_e \geq 10^{12}$ cm$^{-2}$) outstanding successes were achieved in transport studies, e.g., the quantum Hall effect. As an additional technique, methods of optical spectroscopy, such as, e.g., low-temperature (LT) photoluminescence (PL) and photoluminescence excitation (PLE) spectroscopy, offer the advantage of revealing 2DEG-excitation properties. Depending on the carrier density, the dominant excitations are, e.g., neutral excitons, charged excitons (trions), or Fermi-Sea excitations. For $n_e \sim 10^{12}$ cm$^{-2}$, the radiant transitions between photo-created holes and Fermi-Sea electrons were identified by broad, "rectangular" PL spectra (note the flat 2D density of states) starting from the band gap energy $E_0$, with a Fermi edge singularity at their high-energy edge, roughly at $E_0 + E_F$ ($E_F$ is the Fermi energy) [1,2]. These "type-I" spectra were in line with theoretical expectations [3,4]. When considering the lower side of the concentration scale, neutral excitons as well as trions (charged excitons, e.g., two electrons and one hole) are observed, even for nominally undoped

QWs. They manifest themselves in two-line PL spectra ("type-II" spectra), consisting of an exciton recombination line and a second line, redshifted by a few meV, due to radiative recombination inside trions [5,6]. Although there still can be contradictory opinions regarding the existence of free trions [7-10], numerous studies have shown that the exciton-trion picture successfully describes many optical properties of QWs containing low-concentration 2DEGs ($n_e \sim 10^{10}$ cm$^{-2}$) [11]. The trion states can be considered as the first evolutionary stage of the 2DEG response to the presence of the photo-created exciton. For intermediate concentration levels ($n_e \sim 10^{11}$ cm$^{-2}$) between the exciton-trion range and the Fermi-Sea range, no generally accepted understanding of the elementary excitations exists. One interesting theoretical approach originates from Suris, suggesting an effective positive charge (a Fermi-hole) adherent to a conventional trion to form a four-particle bound state [12]. This also can be seen as a repulsion of negative electron background from the negatively charged trion. We call these states "Suris tetrons" (STs) to distinguish them from other four-member quasi-particles [13]. Their experimental identification should be achievable in optical spectroscopy, but the increasing linewidths in this concentration range impede the unambiguous assignment of individual features, especially in optical absorption as proposed by Suris. But in a LT PL, where the STs can be accumulated as an intermediate stage via relaxation from the upper-lying multi-particle states (exciton, trion), they might be identified by proving the original non-trion behavior of the observed spectral features rather than by spectrally resolving the ST. Few experimental studies approached the PL in this concentration range [14-20]. None of them considered ST states, and similar-looking PL spectra were interpreted in different, incompatible ways.

Here, we present an exploration of the possible relevance of Suris´ four-particle excitation to the PL spectra. The spectral-resolution problem is reduced by employing semi-magnetic II-VI QWs, taking advantage of their enhanced spin properties for a sufficient Zeeman-component splitting in a magnetic field, $\vec{B}$. This field is directed in-plane, which sufficiently preserves the orbital wave functions by avoiding Landau level formation. Specifically, we employ PL and spin-flip Raman scattering (SFRS) spectroscopy of (Cd,Mn)Te QWs with intermediate n-doping levels ($n_e \sim 10^{11}$ cm$^{-2}$). Our results, combined with previous experimental findings, can be understood consistently in terms of optically generated *four*-particle Suris-tetron excitations, which thermalize by PL. Details of the thermalized state are deduced from the PL selection rules.

Two MBE-grown modulation-doped QW samples were used, comprising a single (Cd,Mn)Te QW with Cd$_{0.85}$Mg$_{0.15}$Te barriers. QW thicknesses and Mn content were 12 nm and 0.79% (#405A), 15 nm and 0.75% (#303B). Iodine doping was achieved from a $ZnI_2$ source. During growth the substrates were rotated to minimize inhomogeneity. According to a Fermi-velocity analysis [18], the 2DEG concentrations were $2.1 \cdot 10^{11}$ and $2.9 \cdot 10^{11}$ cm$^{-2}$, respectively, so the equation $E_F = \pi \hbar^2 n_e / m_e$ with $m_e \sim 0.1 m_0$ yields $E_F$ ~5 meV and ~7 meV, respectively, which is comparable with the trion binding energy $E_T$ ~3 meV [17]. PL and Raman signals were recorded at 1.5 K in backscattering geometry, using a triple spectrometer and a CCD detector.

Fig.1 (a) and (b) show the zero-field PL spectra, which are typical for similar doping levels in several heterostructure families [15-20]. The two dominating features are the main peak (L2) and a weaker feature at the high-energy side (L1). In addition, a very weak band at the low-energy side (L3) occurs in our samples. Fig.1 (c) shows that L3 is real and, by its pronounced *B*-dependence, is related to the QW states. The lines are significantly broader than the exciton and the trion PL of lower-doped samples of the same quality (~2 meV [16,17]). The PL spectrum of #303B with larger $n_e$ covers a larger energy interval than that of #405A, recalling the spectral evolution with increasing $n_e$ [16-18]; this rapidly increasing PL broadening confirms that our samples fall in the intermediate concentration range of interest.

The *B*-field behavior of the PL features for both samples is depicted in Fig.2. We used orthogonal linear polarizations for excitation and detection and, additionally, rotation of the sample by 90° about the surface-normal axis. Besides non-trivial effects regarding the spin structure of the valence band (VB) [21-23], these configurations yield more precisely the energetic positions of all PL features. All positions strongly depend on the *B*-field with a saturation tendency at high fields due to the giant Zeeman splitting, typical of diluted magnetic semiconductors [24]. For both samples, only the downshifting Zeeman-split branches of L1 and L3 appear, but both branches of L2 (Fig.2). Since the anisotropic heavy-hole (HH) *g*-factor is small for in-plane *B*-fields [25], the shifts of all these features are controlled by the Zeeman energy of the conduction-band (CB) electron, as is concluded from SFRS (see the expected Brillouin-shaped *B*-dependence of the spin-flip energy in Fig.1 (d) and representative SFRS replicas in Fig.3 (a) and (b) near the laser line).

A summary of the optical response of sample #303B in an external *B*-field is presented in Fig.3 (c). In general, #303B and #405A reveal similar *B*-field behavior but different distances between spectral features (cf. Fig.2 (a) and (b)) and a different PL line intensity distribution. For sample #303B, the spectrum in a *B*-field is practically formed by the two Zeeman branches of L2.

An important peculiarity of the PL spectra was observed in sample #303B as the laser energy approached the PL energy from above (Fig.3 (a) and (b); intense sharp lines represent specific laser energies). In both configurations, the upper Zeeman component of L2 abruptly vanishes for a laser energy just above the overall PL collapse and, remarkably, near the hypothetical upper Zeeman branch of L1 (cf. Fig.2 (b)). This supports the interconnected origin of L1 and L2, in contrast with interpretations which ascribe them to different spatial regions of a QW [17,20].

We start the discussion with Fig.1 (a,b). As mentioned, similar PL spectra are observed for various heterostructure families, including CdTe-based QWs. Some authors interpret this shape as merely type-I (which seems natural in view of its broadening with increasing $n_e$) [15,18,19]. In such an interpretation, schematically, the PL spectrum consists of optical transitions involving the HH and (i) $k=0$ electrons (L2) or (ii) electrons at the Fermi edge (L1). The two PL features should then be separated by $E_F$; note for later that transitions at $k=0$ are located at the low-energy side of the PL. The electron-hole Coulomb interaction is often disregarded in this "Fermi-Sea picture", assuming that screening and phase space filling effects make it negligible [15,18].

Light is shed onto the actual PL structure by an important earlier observation [16,17]. By varying $n_e$ in a single sample via secondary illumination, the emergence of the L1, L2 structure was revealed. Remarkably, on increasing $n_e$, the L2 line was found to split off from the L1 line toward low energies [17]. With further increasing $n_e$, L2 still downshifts and grows, while L1 keeps its position near the trion energy (which was reliably identified for samples with low $n_e$ [16,17]). Teran et al. [17] ascribed L2 to the Fermi-Sea transitions affected by photo-created holes, and L1 to the trion-like recombination, whose exact origin was to be clarified.

Recently, a different concept of the PL spectrum was put forward [20], in which L1 is still due to the trion optical emission (which is an attractive suggestion since, on increasing $n_e$, L1 inherits the trion energy position). L2 was ascribed to a trion emitting a plasmon together with the photon [20], whose radiation is thereby red-shifted by the characteristic plasmon energy. This interpretation is compatible with Ref. [17], and the increasing distance between L1 and L2 was naturally explained. The models of both Ref. [20] and Ref. [17] invoke a certain concentration inhomogeneity for allowing the simultaneous observation of L1 and L2 in one system. Interestingly, in the picture of Ref. [20], transitions at $k=0$ (or rather, the trion transitions closest to them) are located at the high-energy side of the PL. This highlights the fact that the difference between the two interpretations is not merely terminological.

The behavior of the L1, L2 lines shown in Refs. [16,17] is difficult to understand within the "Fermi-Sea picture", even when accounting for the carrier-induced band gap renormalization (BGR) [26], because the emission is energetically misplaced with respect to the exciton transition energy in the low-$n_e$ limit, $E_X$. The calculated BGR downshift of the CB minimum for $2 \cdot 10^{11}$ cm$^{-2}$ is as low as ~14 meV [17] (or even ~8 meV [27]), while the measured exciton binding energy in the low-$n_e$ limit is 18 meV [17]. So, if the emission would originate from the VB hole and the Fermi-Sea electrons, and assuming the Coulomb attraction fully screened at such concentrations, the PL low-energy onset ($k=0$) would be at 4, or even at 10 meV above $E_X$. However, experimentally, the entire PL spectrum is well below $E_X$ (see Fig.1 in Ref.[17]). This probably reveals the persisting relevance of Coulomb interaction for these concentrations.

On the other hand, our observations regarding L1 and L2 do not seem compatible either with the alternative, trion-related interpretations of Refs. [17,20]. Although L1 inherits the trion energetic position, it does not show both Zeeman branches, as the trion PL in the Voigt configuration is known to do [22,28]. Both branches occur for L2, not for L1 (Fig.2). In view of these inconsistencies, we suggest a new interpretation of the PL spectrum, based on the Suris tetron concept. This four-member quasi-particle was obtained in Ref.[12] as a sharp peak in the calculated light absorption spectrum of a 2DEG in a QW and was identified as a trion bound to a hole in the Fermi Sea of the CB electrons (a Fermi-hole, FH) [29]. The FH appears in the Fermi Sea when the photogenerated virtual electron-hole pair captures an extra CB electron to form the trion. STs have not been observed in absorption spectra, presumably because their energy separation from the trion states is too small compared to typical linewidths. However, being the lowest bound states in the system, the STs should play an important role in LT PL. It is a reasonable assumption that, some time after creation, the majority of optical excitations in the 2DEG form STs. We suppose that the ST is the initial state of the optical transitions forming the PL, while various possible final states give the various features in the PL spectra (Fig.3 (d)).

Within this interpretation, L1 results from the "double recombination" of the ST: one electron (E1) recombines with the HH, while the opposite-spin electron (E2) recombines with the FH and leaves behind the unperturbed Fermi Sea (Fig.3 (d)). This explains the *B*-field shift of L1 (the FH has the same $g$-factor as the CB electron), as well as the absence of the upper Zeeman branch of L1 (the initial-state splitting exceeds $k_B T$). The inheritance by L1 of the energetic position of the trion with increasing $n_e$ then follows naturally.

The usual recombination E1-HH is also still possible and leaves behind the E2-FH pair. Such a pair is an excitation of the electron density (probably related to a plasmon [20]); here we need to consider only the pair spin state. The two particles related to one band should possess a significant exchange interaction and will form a singlet or a triplet state; the recombination scheme thus resembles that of a doubly charged exciton $X^{2-}$ in a quantum dot [30]. Within our model, the transitions responsible for L2 yield a triplet final state (Fig.3 (d)). This ensures comparable intensities of both Zeeman branches, as observed. The reason is analogous to that well known for trion PL: this Zeeman splitting originates from the transition´s *final* state, where a thermal population factor is irrelevant. Transitions to the upper triplet state are forbidden by selection rules, because only the lower initial tetron state is occupied.

Finally, L3 can be ascribed to transitions into the singlet final state. This is consistent with the presence of only the lower Zeeman branch of L3. Note also the nearly equal spacing between L2 and L3 in our two samples (6 and 6.5 meV); if ascribed to the singlet-triplet separation, it agrees quite well with the few-tens meV scale of the intraband electron-electron exchange interaction $E_{exch}$ reported for QDs [30,31]. The spacing between L1 and L2 is noticeably different for the two samples with different $n_e$ (2.8 and 5.1 meV), in agreement with calculated 2D-plasmon energies $E_{2Dp}$ [20].

The disappearance of the upper Zeeman component of L2 at some threshold excitation energy $E_{th}$ (Fig.3) means that above- and below-threshold excitations prepare the emitting system differently; the initial state of the PL transitions is not fully relaxed at least in one case. In terms of the ST as the initial state, possible candidates for not being relaxed are the HH and/or FH spins (E1 and E2 form a singlet and are ruled out). By the absence of the upper Zeeman branches of L1 and L3 (Fig.2), combined with the assignment of these lines in Fig.3 (d), we conclude that the FH populates its low-energy spin sublevel $|\downarrow_{FH}\rangle$ before the photon emission, i.e., the initial-state FHs are spin-relaxed. We are led to the assumption that the HHs are not relaxed and their spin state is conserved from creation until recombination (this should be valid at least for near-threshold ST quasi-resonant excitation).

Let us consider the ST optical selection rules. Suppose the incoming light creates a virtual E2-HH pair that combines with a virtual E1-FH pair excitation of the Fermi Sea, to result in the resonant ST formation. In an in-plane *B*-field, the ST state is either (i) $\frac{1}{\sqrt{2}}(|\uparrow_{E1}\downarrow_{E2}\rangle - |\downarrow_{E1}\uparrow_{E2}\rangle)|\uparrow_{FH}\rangle|\Uparrow_{HH}\rangle$ or (ii) $\frac{1}{\sqrt{2}}(|\uparrow_{E1}\downarrow_{E2}\rangle - |\downarrow_{E1}\uparrow_{E2}\rangle)|\downarrow_{FH}\rangle|\Uparrow_{HH}\rangle$; the arrows denote projections onto the *B*-field direction, pointing to the particle's high-energy (↑) or low-energy (↓) Zeeman state. Case (i) corresponds to excitation of the upper-, case (ii) – of the lower Zeeman branch of the ST (Fig.3 (d)), and one can take an arbitrary projection of the HH pseudospin. If state (i) is excited, it subsequently relaxes and emits from state (ii). Observation in crossed polarizations, in combination with HH spin state conservation, means that while HH is generated with the electron E2, it annihilates with the opposite-spin electron E1 [21]. Suppose the $|\downarrow_{E2}\Uparrow_{HH}\rangle$ pair is excited, which is completed to the full state (i) by the $|\uparrow_{E1}\uparrow_{FH}\rangle$ pair; after the FH spin flip leading to state (ii), the subsequent radiative transition takes the system exactly to the lowest triplet final state $|\downarrow_{E2}\downarrow_{FH}\rangle$. This process manifests itself as the upper Zeeman branch of L2. Remarkably, it vanishes when the incoming photon energy becomes insufficient to excite the upper ST branch, the (i) state ($E_{th}$ in Fig. 3(d)). At the same time, the incoming $|\uparrow_{E2}\Downarrow_{HH}\rangle$ pair is completed by the $|\downarrow_{E1}\downarrow_{FH}\rangle$ pair, to result in the direct (ii)-type state formation. The radiative transition now results in the central triplet final state, $\frac{1}{\sqrt{2}}(|\uparrow_{E2}\downarrow_{FH}\rangle + |\downarrow_{E2}\uparrow_{FH}\rangle)$. This process induces the lower Zeeman branch of L2, and works as long as the lower ST branch can be excited (Fig.3 (d)). This explains the different excitation edges for both branches of L2.

Note that the above explanation relies on the assumption that non-spin-conserving pair excitations of the Fermi Sea ($|\uparrow_{E1}\uparrow_{FH}\rangle$ or $|\downarrow_{E1}\downarrow_{FH}\rangle$) trap the E2-HH pairs more efficiently than the spin-conserving ones ($|\downarrow_{E1}\uparrow_{FH}\rangle$ or $|\uparrow_{E1}\downarrow_{FH}\rangle$). This might be related to the longer lifetimes of the non-spin-conserving virtual excitations. To clarify such points, further development of the theory of ST states is necessary.

In summary, our PL study of QWs with intermediate 2DEG concentrations ($n_e \sim 10^{11}$ cm$^{-2}$) revealed a characteristic *B*-field behaviour. Based on these observations and earlier findings by other authors, we report a new understanding of the PL spectrum in terms of four-particle 2DEG-excitations (Suris tetrons). The PL features are attributed to radiative transitions from these quasi-particles to different final states. Within this context, for (Cd,Mn)Te QWs the PL develops from narrow-line exciton-trion spectra to broad-band Fermi-Sea recombination spectra via few-particle states [32] rather than via screening and band gap renormalization effects.

We appreciate valuable discussions with R.A. Suris, A.A. Klochikhin and K.V. Kavokin. This work was partially supported by RFBR (project 13-02-00316-a), by Samsung 2011 GRO Project and by Russian Ministry of Education and Science (contract No. 11.G34.31.0067 with SPbSU). AK gratefully acknowledges support from Dmitry Zimin "Dynasty" Foundation.

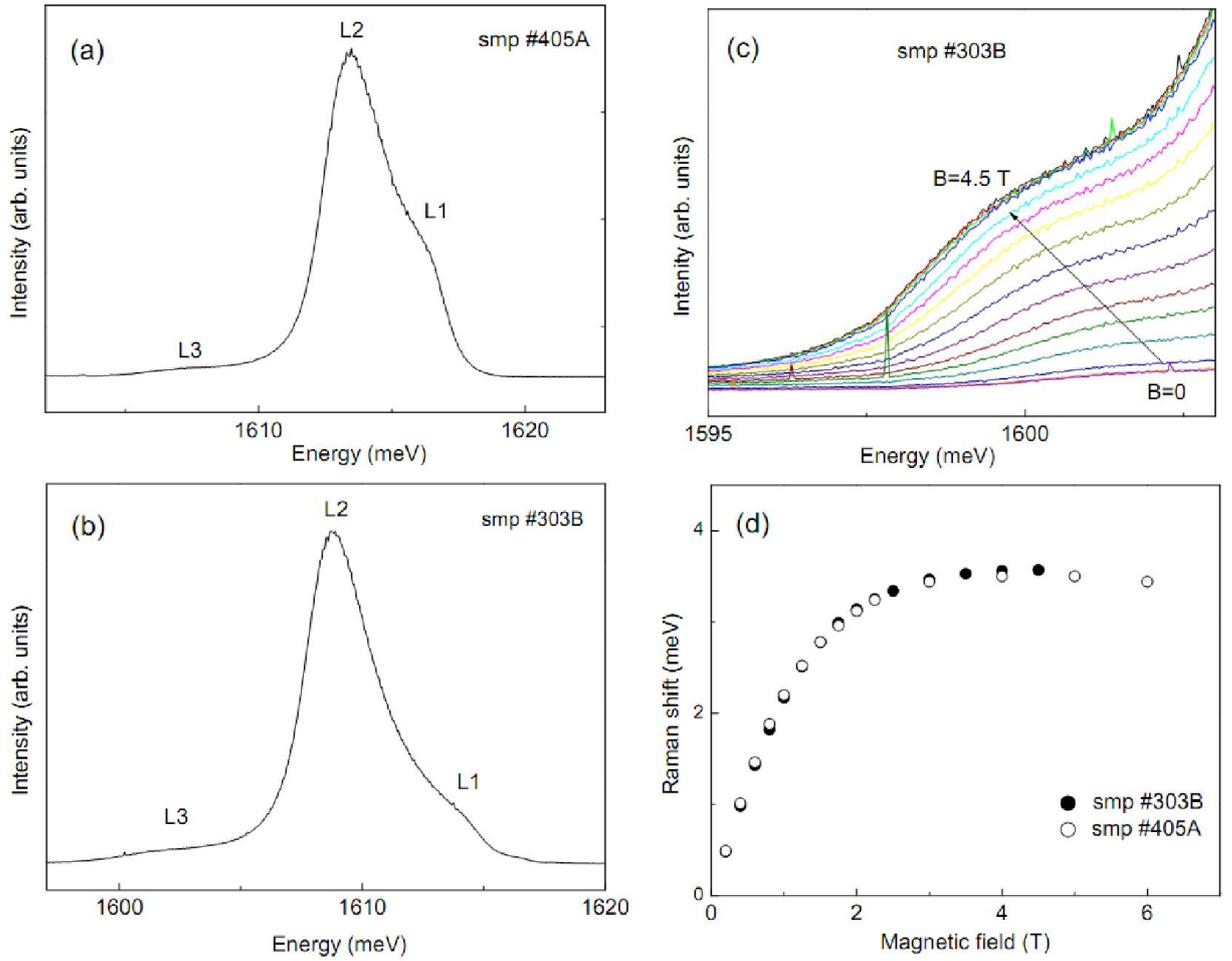

Fig.1 Zero-field PL spectra of samples (a) #405A (b) #303B at laser excitation energies well above the QW emission; notations for the PL features (L1, L2, L3) are indicated. Panel (c) demonstrates the L3 line behavior in #303B vs. the $B$-field, supporting its attribution to the QW layer. Panel (d) shows the $B$-field dependence of the SFRS Raman shift in both samples, a direct measure of the Zeeman splitting of the $c$-band electron states. Pump power 24 µW everywhere.

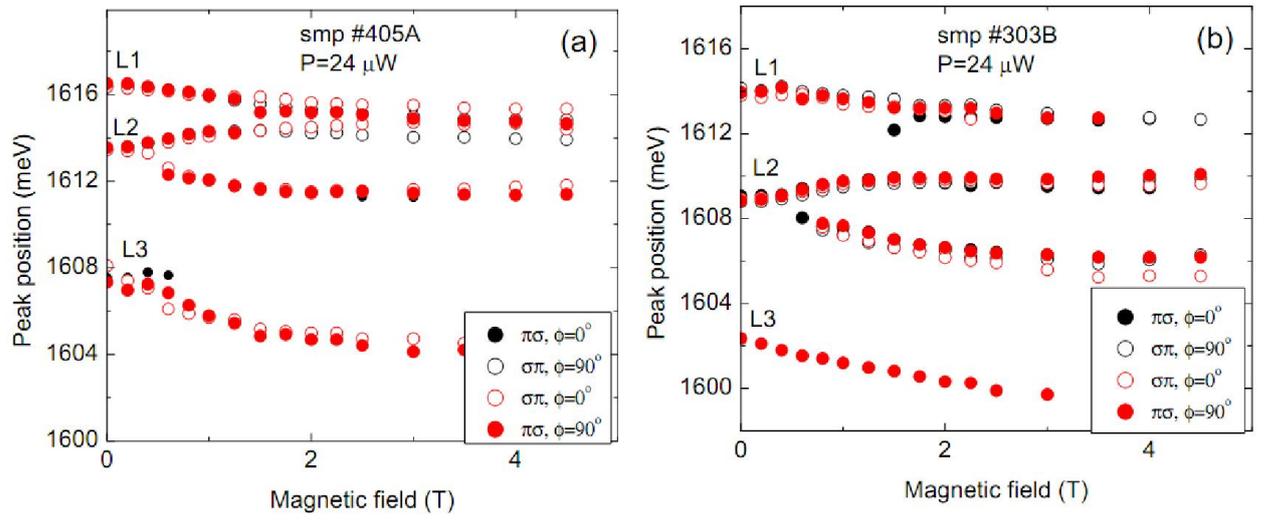

Fig.2 *B*-field dependence of positions of the PL features in the Voigt configuration recorded at various polarization setups (see text) for samples #405A (a) and #303B (b).

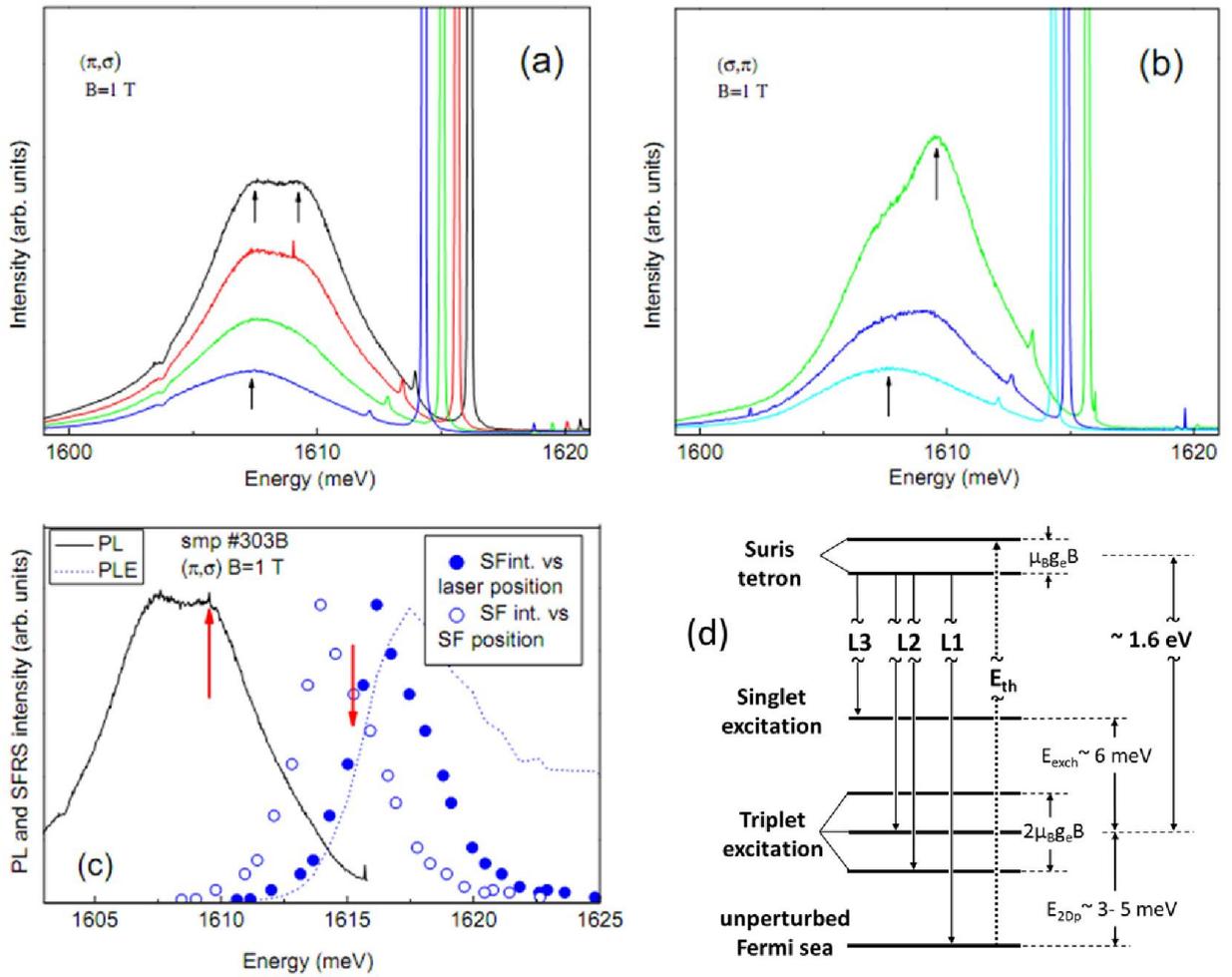

Fig.3 Optical response of #303B at $B = 1$ Tesla, laser power $P = 8\,\mu W$ (a-c). Panels (a) and (b) demonstrate the disappearance, below a threshold excitation energy $E_{th}$, of the upper Zeeman branch of the L2 line. Panel (c) shows a representative PL spectrum (solid line), the PLE spectrum (dotted line) and the resonance profile of the SFRS, plotted twice for convenience: as functions of the laser-line and the Raman-line positions, respectively (circles). The upward arrow points to the vanishing L2 branch, the downward arrow – to (approximately) the threshold laser energy $E_{th}$ for which it vanishes. Panel (d) depicts the energy scheme of the radiative transitions connecting the Suris tetron and various states of the Fermi Sea. The assignment of the experimental PL features is indicated, and the threshold excitation energy $E_{th}$ for the collapse of the upper L2 branch in PL is indicated by the vertical dotted arrow.